\documentclass[
twocolumn,
hf,
]{ceurart}

\usepackage{listings}
\usepackage{tabularx}
\usepackage{hyperref}
\usepackage{lscape}
\usepackage{hvfloat}
\begin{document}

\copyrightyear{2022}
\copyrightclause{Copyright for this paper by its authors.
  Use permitted under Creative Commons License Attribution 4.0
  International (CC BY 4.0).}

\conference{JCDL'22: ULITE-ws, Understanding LIterature references in academic full TExt,
  June 24--06, 2022, Cologne, Germany}

\title{How to structure citations data and bibliographic metadata in the OpenCitations accepted format}

\author[1]{Arcangelo Massari}[%
orcid=0000-0002-8420-0696,
email=arcangelo.massari@unibo.it,
]
\author[1]{Ivan Heibi}[%
orcid=0000-0001-5366-5194,
email=ivan.heibi2@unibo.it,
]
\address[1]{Research Centre for Open Scholarly Metadata, Department of Classical Philology and Italian Studies, University
of Bologna, Bologna, Italy}


\begin{abstract}
The OpenCitations organization is working on ingesting citation data and bibliographic metadata directly provided by the community (e.g., scholars and publishers). The aim is to improve the general coverage of open citations, which is still far from being complete, and use the provided metadata to enrich the characterization of the citing and cited entities. This paper illustrates how the citation data and bibliographic metadata should be structured to comply with the OpenCitations accepted format.
\end{abstract}


\maketitle

\section{Introduction}

The Declaration on Research Assessment \cite{cagan_san_2013}, the Leiden Manifesto for Research Metrics \cite{hicks_bibliometrics_2015}, and the Initiative for Open Citations (I4OC, \url{https://i4oc.org/}) have successfully convinced almost all major academic publishers to release their publication reference lists. To date, more than 1.2 billion citations are available through the Crossref REST API \cite{hendricks_crossref_2020} and distributed by OpenCitations \cite{peroni_opencitations_2020} as structured, separated from the original bibliographic source and under the CC0 license \cite{peroni_open_2018}.

Nevertheless, the coverage of open citations is still far from complete \cite{martin-martin_coverage_2021}. On the one hand, some publishers have not yet made their citations public. On the other hand, many citations are lost because they are only present in unstructured format within PDF files, especially in social sciences.

OpenCitations is working on ingesting citations and bibliographic metadata directly coming from the community (e.g., scholars and publishers). In this way, projects like EXCITE \cite{hosseini2019excite} - aimed at extracting citations from PDFs - could significantly contribute to increasing the data coverage.

The following section illustrates how to structure the citation data and bibliographic metadata in the accepted format of OpenCitations. We conclude this paper with a description of the upcoming future related works.

\section{Metadata and citations}\label{body}

OpenCitations manages and processes two different CSV files to separately characterize the ingested documents, one containing their metadata (META-CSV), and a second one holding their citations (CITS-CSV). On this section we discuss how these files should be structured and defined before providing them to OpenCitations. The discussion presented in this section is based on a more exhaustive documentation  \cite{massari_arcangelo_2022_6597141}.

CSV files are logically structured as tables. In META-CSV each document (row), is characterised by 11 attributes (columns): 

\begin{itemize}
  \item \textbf{id}. the ID(s) of the corresponding document. A document can have more than one ID, each ID is defined by its type (using an acronym) and value. Multiple IDs must be separated using single white space, as follow:
  \begin{center}ID abbreviation + ``:'' + ID value\end{center}
  For example ``doi:10.3233/ds-170012'' indicates a DOI identifier having the value ``10.3233/ds-170012''.
  \item \textbf{title}. a textual value to express the title of the document.
  \item \textbf{author} and \textbf{editor}. Data regarding the authors and the editors of the document. Each character (author/editor) is defined by several attributes, e.g., his family name or ID. Multiple characters are separated by a semicolon followed by a white space character. Generally, the definition of an actor follows this structure:
  \begin{center}Family Name + ``,'' + `` '' + Given Name + `` '' + ``['' + IDs + ``]''\end{center}
  The IDs of the authors/editors are specified in square brackets and follow the format used for the ``id'' attribute.
  \begin{center}e.g. ``Peroni, Silvio [orcid:0000-0003-0530-4305]''\end{center}
  In case of no IDs, the square brackets are omitted from the character description either. The given name is not mandatory, however, the description of the character should still contain a comma to indicate such absence (e.g. ``Peroni, [orcid:0000-0003-0530- 4305]'')
  \item \textbf{pub\_date}. the date of publication of the document. The date is defined according to \href{https://www.iso.org/iso-8601-date-and-time-format.html}{ISO 86014}\cite{wolf_date_nodate}, the ISO standard for ``Representation of dates and times'':
  \begin{center}YYYY-MM-DD\end{center}
  It is mandatory to specify at least the publication year. The values of the month and day are not required. However, if the day is specified, the month must be specified as well.
  \item \textbf{venue}. data regarding the venue of the document. For example, if the document is a journal article, the venue defines the journal where the document has been published. Each venue is described as follows:
  \begin{center}Venue Title + `` '' + ``['' + IDs + ``]''\end{center}
  The IDs of a venue are described using the same format used previously. In case of no identifiers, the square brackets are omitted.
  \item \textbf{volume} and \textbf{issue}. these values are required only if the document is contained in a journal volume or a journal issue.
  \item \textbf{page}. the page range of the corresponding document, defined through the specification of the first and the last page, divided by a hyphen ``-''.
  \item \textbf{type}. a textual value to identify the document. This value is taken from the list of the currently supported bibliographic resource types: book, book chapter, book part, book section, book series, book set, book track, component, dataset (or data file), dissertation, edited book, journal, journal article, journal issue, journal volume, monograph, other, peer review, posted content (or web content), proceedings, proceedings article, proceedings series, reference book, reference entry, report, report series, standard, and standard series.
  \item \textbf{publisher}. the publisher name of the corresponding document. To define a publisher we apply the same format used in the definition of the \texttt{venue}.
\end{itemize}

If the resource identifier is specified in the ``id'' field, all the other fields are optional.
Conversely, if the ``id'' field is empty, there are mandatory fields that vary depending on the resource type:

\begin{itemize}
    \item The fields ``title'', ``pub\_date'', and ``author'' (or ``editor'') are mandatory for the resources of type book, dataset (or data file),  dissertation, edited book, journal article, monograph, other, peer review, posted content (or web content), proceedings article, report, and reference book. Moreover, this information is compulsory if the "type" field is empty.
    \item The ``title'' and ``venue'' fields are required for the resources of type book chapter, book part, book section, book track, component, and reference entry.
    \item Only the ``title'' field is required for the resources of type book series, book set, journal, proceedings, proceedings series, report series, standard, and standard series.
    \item Regarding the resources of journal volume type, the fields ``venue'' and ``volume'', or ``venue'' and ``title'', are mandatory. Conversely, as for resources of journal issue type, the fields ``venue'' and ``issue'', or ``venue'' and ``title'', are mandatory.
\end{itemize}

Table \ref{metadata_csv} shows an example of a well-formed META-CSV representation. The table contains a small sample of ten documents (rows) and their corresponding attributes (columns).

On the other hand, in the CITS-CSV each entity (row) represents a citation. A citation is characterised by 4 attributes (columns): \textbf{citing\_id}, \textbf{citing\_publication\_date}, \textbf{cited\_id}, and \textbf{cited\_publication\_date}. The \texttt{citing\_id} and \texttt{cited\_id} values represent the identifiers of the citing and cited document, respectively. These values are both mandatory, and they are structured following the same scheme used for \textbf{id} definition in META-CSV. The \textbf{citing\_publication\_date} and \textbf{cited\_publication\_date} represent the date of publication of the citing and cited document, respectively. Both these values are optional, and  follow the same structural scheme used for \texttt{pub\_date} definition in META-CSV. 

Table \ref{citations_csv} shows an example of a well-formed CITS-CSV representation. The table contains a small sample of ten different citations (rows) and their corresponding attributes (columns).

\section{Discussion and conclusion}\label{conclusion}

This paper described how to define well-formed CSV files storing citations and metadata of bibliographic resources, ready to be provided and later processed by OpenCitations.

The ingestion of bibliographic metadata will be possible starting from the release of OpenCitations Meta (OC-Meta), expected by the end of 2022. OC-Meta will store bibliographic metadata for the documents involved (as citing or cited entities) in OpenCitations citation indexes.

The ingestion of the citations is possible thanks to CROCI, the Crowdsourced Open Citations Index, which allows individuals identified by ORCIDs to deposit the citation data that they
have legal right to submit \cite{DBLP:journals/corr/abs-1902-02534}. Citation data are submitted to either Figshare (\url{https://figshare.com}) or Zenodo (\url{https://zenodo.org}), accompanied by the ORCID of the contributor. Aftwerwards, the submitter can inform OpenCitations using the GitHub issue tracker on the CROCI repository (\url{https://github.com/opencitations/croci/issues}).

Future works include implementing an interface that simplifies and automates the entire publication process via CROCI, also providing input data validation and modification suggestions. 

Moreover, CROCI currently handles only DOI-to-DOI citations. The upcoming plan is to let CROCI manage also any-to-any citations.
\begin{acknowledgments}
    This work was funded from the European Union's Horizon 2020 research and innovation program under grant agreement No 101017452 (OpenAIRE-Nexus Project). We want to thank Silvio Peroni for supervising the entire work on OpenCitations, Philipp Mayr-Schlegel and Ahsan Shahid for the feedback on the documentation from which this demo paper is drawn, and Davide Brambilla for the valuable insights about CROCI and its future developments.
\end{acknowledgments}

\bibliography{bibliography}

\newpage
\appendix

\section{Appendix}


\hvFloat[
floatPos=p,
rotAngle=90,
capPos=top,
objectPos=center
]{table}{
\resizebox{\textheight}{!}{
\begin{tabular}{lllllllllll}
\hline
\textbf{id} & \textbf{title} & \textbf{author} & \textbf{pub\_date} & \textbf{venue} & \textbf{volume} & \textbf{issue} & \textbf{page} & \textbf{type} & \textbf{publisher} & \textbf{editor} \\ \hline
doi:10.1007/978-3-030-00668-6\_8 & The SPAR Ontologies & \begin{tabular}[c]{@{}l@{}}Peroni, Silvio {[}orcid:0000-0003-0530-4305{]}; \\ Shotton, David {[}orcid:0000-0001-5506-523X{]}\end{tabular} & 2018 & \begin{tabular}[c]{@{}l@{}}17th ISWC \\ {[}doi:10.1007/978-3-030-00668-6{]}\end{tabular} &  &  & 119-136 & book chapter & \begin{tabular}[c]{@{}l@{}}Springer International \\ Publishing \\ {[}crossref:297{]}\end{tabular} &  \\ \hline
doi:10.3233/DS-170012 & Automating semantic publishing & Peroni, Silvio {[}orcid:0000-0003-0530-4305{]} & 2017 & \begin{tabular}[c]{@{}l@{}}Data Science \\ {[}issn:2451-8484 issn:2451-8492{]}\end{tabular} & 1 & 1-2 & 155-173 & journal article & \begin{tabular}[c]{@{}l@{}}IOS Press \\ {[}crossref:7437{]}\end{tabular} &  \\ \hline
\begin{tabular}[c]{@{}l@{}}doi:10.1007/978-3-476-00160-3 \\ isbn:9783476021144 \\ isbn:9783476001603\end{tabular} & Literatur &  & 2005 &  &  &  &  & book & \begin{tabular}[c]{@{}l@{}}Springer Science and \\ Business Media LLC \\ {[}crossref:297{]}\end{tabular} & Gfrereis, Heike \\ \hline
\begin{tabular}[c]{@{}l@{}}doi:10.1057/9780230316645 \\ isbn:9780230276604 \\ isbn:9780230316645\end{tabular} & New Waves in Philosophy of Law &  & 2011 &  &  &  &  & book & \begin{tabular}[c]{@{}l@{}}Springer Science and \\ Business Media LLC \\ {[}crossref:297{]}\end{tabular} & Mar, Maksymilian Del \\ \hline
\begin{tabular}[c]{@{}l@{}}doi:10.4324/9781003115830 \\ isbn:9781003115830\end{tabular} & Governing Savages & Markus, Andrew & 2020-7-31 &  &  &  &  & book & \begin{tabular}[c]{@{}l@{}}Informa UK Limited \\ {[}crossref:301{]}\end{tabular} &  \\ \hline
\begin{tabular}[c]{@{}l@{}}doi:10.1515/9781503600836 \\ isbn:9781503600836\end{tabular} & Newsworthy & Barbas, Samantha & 2020-6-24 &  &  &  &  & book & \begin{tabular}[c]{@{}l@{}}Walter de Gruyter GmbH \\ {[}crossref:374{]}\end{tabular} &  \\ \hline
doi:10.1134/s0018151x17020055 & \begin{tabular}[c]{@{}l@{}}On the theory of \\ convection of electrons in metals\end{tabular} & Gladkov, S. O. & 2017-5 & \begin{tabular}[c]{@{}l@{}}High Temperature \\ {[}issn:0018-151X issn:1608-3156{]}\end{tabular} & 55 & 3 & 321-325 & journal article & \begin{tabular}[c]{@{}l@{}}Pleiades Publishing Ltd \\ {[}crossref:137{]}\end{tabular} &  \\ \hline
doi:10.1134/s0018151x17050029 & Stability of boiling shock & Avdeev, A. A. & 2017-9 & \begin{tabular}[c]{@{}l@{}}High Temperature \\ {[}issn:0018-151X issn:1608-3156{]}\end{tabular} & 55 & 5 & 753-760 & journal article & \begin{tabular}[c]{@{}l@{}}Pleiades Publishing Ltd \\ {[}crossref:137{]}\end{tabular} &  \\ \hline
doi:10.1134/s0018151x17050224 & \begin{tabular}[c]{@{}l@{}}The high-temperature \\ and radiative effect on concrete\end{tabular} & Zhakin, A. I. & 2017-9 & \begin{tabular}[c]{@{}l@{}}High Temperature \\ {[}issn:0018-151X issn:1608-3156{]}\end{tabular} & 55 & 5 & 767-776 & journal article & \begin{tabular}[c]{@{}l@{}}Pleiades Publishing Ltd \\ {[}crossref:137{]}\end{tabular} &  \\ \hline
doi:10.1134/s0018151x18010169 & \begin{tabular}[c]{@{}l@{}}Relaxation of Rayleigh \\ and Lorentz Gases in Shock Waves\end{tabular} & Skrebkov, O. V. & 2018-1 & \begin{tabular}[c]{@{}l@{}}High Temperature \\ {[}issn:0018-151X issn:1608-3156{]}\end{tabular} & 56 & 1 & 77-83 & journal article & \begin{tabular}[c]{@{}l@{}}Pleiades Publishing Ltd \\ {[}crossref:137{]}\end{tabular} &  \\ \hline
\end{tabular}
}}{A sample of ten documents characterized by their corresponding metadata attributes}{metadata_csv}

\newpage
\hvFloat[
floatPos=p,
rotAngle=90,
capPos=top,
objectPos=center
]{table}{
\begin{tabular}{llll}
\hline
\textbf{citing\_id} & \textbf{citing\_publication\_date} & \textbf{cited\_id} & \textbf{cited\_publication\_date} \\ \hline
doi:10.1016/j.websem.2012.08.001 & 2012-12 & doi:10.1087/2009202 & 2009-04-01 \\ \hline
doi:10.1016/j.websem.2012.08.001 & 2012-12 & doi:10.1371/journal.pcbi.1000361 &  \\ \hline
doi:10.1016/j.websem.2012.08.001 & 2012-12 & doi:10.1007/978-3-642-33876-2\_35 & 2012 \\ \hline
doi:10.1016/j.websem.2012.08.001 & 2012-12 & doi:10.1186/2041-1480-1-S1-S6 & 2010-06-22 \\ \hline
doi:10.1016/j.websem.2012.08.001 & 2012-12 & doi:10.1145/945645.945664 & 2003-10-23 \\ \hline
pmid:23636598 & 2013 & pmid:19151427 & 2005 \\ \hline
pmid:23636598 & 2013 & pmid:19782561 & 2008-10 \\ \hline
pmid:23636598 &  & pmid:18686754 & 2012-09-05 \\ \hline
pmid:23636598 & 2013 & pmid:15890079 & 2009-07-15 \\ \hline
pmid:23636598 & 2013 & pmid:18191757 &  \\ \hline
\end{tabular}
}{A sample of ten citations characterized by their related attributes}{citations_csv}
\end{document}